\pdfoutput=1
\documentclass[conference]{IEEEtran}
\usepackage{etex}
\usepackage{amssymb}
\usepackage{times}
\usepackage{bm}
\usepackage{subcaption}
\usepackage{amsmath}
\usepackage{graphicx}
\usepackage{amssymb}
\usepackage{amstext}
\usepackage{latexsym}
\usepackage{pifont}
\usepackage{color,soul}
\usepackage{ifthen}
\usepackage{multirow}
\usepackage[linesnumbered,ruled]{algorithm2e}
\usepackage{verbatim}
\usepackage{array,tabularx}
\usepackage{accents}
 \usepackage{cite}
 \usepackage{arydshln}
\usepackage{hyperref}
\usepackage{url,amssymb,epsfig,times,mathrsfs,algorithmic,latexsym,xspace,fancyhdr,wrapfig, xcolor,multirow,amsthm}
\usepackage{caption}
\usepackage{soul}
\usepackage[inline]{enumitem}
\usepackage{arydshln}
\usepackage{enumitem}
\usepackage[makeroom]{cancel}
\usepackage{mathtools}
\usepackage{dblfloatfix}
\usepackage[utf8]{inputenc}
\usepackage{dsfont}

\newcommand{\Mod}[1]{\ \mathrm{mod}\ #1}

\hypersetup{nolinks=true}
\usepackage{tikz}
\usetikzlibrary{shapes.geometric}
\usetikzlibrary{shapes,snakes,patterns}
\usetikzlibrary{arrows,positioning,automata,calc}
\usetikzlibrary{plotmarks}
\tikzset{
    int/.style={
           rectangle,
           rounded corners,
           draw=black, thin, fill=black!20,
           minimum height=2em,
           inner sep=2pt,
           text centered,
           },
}

\newtheorem{proposition}{Proposition}%[section]
\newtheorem{example}{Example}%[section]
\IEEEoverridecommandlockouts

\renewcommand{\baselinestretch}{1}
\pdfoutput=1

\begin{document}
\pdfoutput=1
\allowdisplaybreaks
\newlength\figureheight
\newlength\figurewidth

\renewcommand{\baselinestretch}{.95}

\title{Codes Correcting Burst and Arbitrary Erasures for Reliable and Low-Latency Communication}
\author{
\IEEEauthorblockN{Serge Kas Hanna\IEEEauthorrefmark{1}, Zhiyuan Tan\IEEEauthorrefmark{3}, Wen Xu\IEEEauthorrefmark{4}, and  Antonia Wachter-Zeh\IEEEauthorrefmark{2}}
\IEEEauthorblockA{
\IEEEauthorrefmark{1} Department of Mathematics and Systems Analysis, School of Science, Aalto University, Finland \\
\IEEEauthorrefmark{2} Institute for Communications Engineering, Technical University of Munich, Germany \\
\IEEEauthorrefmark{3}  Huawei Technologies Co., Ltd., Shanghai, China \\
\IEEEauthorrefmark{4} Munich Research Center, Huawei Technologies Duesseldorf GmbH, Munich, Germany \\
Emails: serge.kashanna@aalto.fi, tanzhiyuan@huawei.com, wen.dr.xu@huawei.com, antonia.wachter-zeh@tum.de
}
\vspace{-0.8cm}
}

\maketitle
\begin{abstract} 
Motivated by modern network communication applications which require low latency, we study codes that correct erasures with low decoding delay. We provide a simple explicit construction that yields convolutional codes that can correct both burst and arbitrary erasures under a maximum decoding delay constraint $T$. Our proposed code has efficient encoding/decoding algorithms and requires a field size that is linear in $T$. We study the performance of our code over the Gilbert-Elliot channel; our simulation results show significant performance gains over low-delay codes existing in the literature.
\end{abstract}

\section{Introduction}
In network communication, latency is defined as the time elapsed between the transmission of a packet from the source and its recovery at the receiver. Ensuring reliable communication with low latency is the main challenge in designing next-generation communication systems. To this end, we focus on studying error-correcting codes that ensure reliability in the presence of low decoding delay constraints. Our motivation stems from several modern applications such as Extended Reality (XR), Virtual Reality (VR), and Cloud Gaming. %In such applications, having reliable, low-latency communication is crucial for real-time, high-resolution video streaming. 

In this work, we focus on forward error correction (FEC). FEC codes for correcting erasures (i.e., packet drops) with low decoding delay were first investigated by Martinian {\em et al.} in~\cite{Martinian2004, Martinian2007}. The works in \cite{Martinian2004, Martinian2007} focused on designing codes that correct a burst of $B$ erasures with a decoding delay that is at most $T$, and showed that the optimal code rate for this setting is $R=\frac{T}{T+B}$. In~\cite{Martinian2007}, Martinian and Trott presented an explicit rate-optimal code that can be constructed for all values of $B, T$, and $R$, with a linear field size $q=O(T)$. In~\cite{Badr2017}, the authors generalized the study on codes with low decoding delay by introducing and studying a channel model that considers both burst and arbitrary erasures. This channel model was referred to as the sliding-window erasure channel, denoted by $C(N, B, W)$, where in any sliding window of size $W$, the channel can introduce either an erasure burst of maximum length $B$, or else, up to $N$ arbitrary erasures. The authors in \cite{Badr2017} showed that, without loss of generality, the channel parameter $W$ could be set to $W=T+1$, and the optimal code rate for this channel is~$R=\frac{T+1-N}{T+1-N+B}$. %The scope of the results in~\cite{Martinian2004, Martinian2007} is limited to the case of a burst erasure, i.e., the special case of  $C(N=1, B, W)$. 

The term {\em streaming codes} is widely used in the literature to refer to codes that can correct {\em all} erasure patterns generated by the $C(N, B, W)$ sliding-window channel model. A series of works focused on designing rate-optimal streaming codes, e.g.,~\cite{Badr2017,Kumar2018, Fong2019l, dudzicz2019explicit, Rakumar2020, Domanovitz2020, Kumar2020IT, Kumar2020L,fong2020optimal,badr2017fec,badr2011diversity}. These codes have different field size requirements in terms of $T$, ranging from exponential $q=O(e^T)$ to linear $q=O(T)$. Existing streaming codes with linear field sizes can be constructed only for certain values of $N, B,$ and  $T$, e.g., \cite{Rakumar2020, Kumar2020L}. The state-of-the-art optimal streaming code that can be constructed for {\em all} values of the parameters $N, B, T$, is the one that was recently presented by Domanovitz {\em et al.}~\cite{Domanovitz2020}, which requires a quadratic field size $q=O(T^2)$. Some works also considered variations of the $C(N, B, W)$ sliding-window channel, the authors in~\cite{badr2015streaming} proposed a code with cubic field size $q=O(T^3)$ that can correct most erasures when a window of size $W$ contains a {\em single} arbitrary erasure {\em in addition} to a burst erasure of maximum size $B$. 

The goal of introducing sliding-window channels such as the $C(N, B, W)$ channel in~\cite{Badr2017} was to provide a theoretically tractable approximation of the Gilbert-Elliot (GE) channel\cite{Gilbert, Elliot}, which is a random channel with memory that is commonly used to model packet erasures over networks. However, the accuracy of this approximation heavily depends on the values of $N,B,T$, and the GE channel parameters. Namely, some realizations of the GE channel may include both burst and multiple arbitrary erasures occurring in the same window, and such cases are not covered by the sliding-window erasure channels studied in~\cite{Badr2017} and~\cite{badr2015streaming}. These realizations could have a relatively high probability of occurrence depending on the values of the system parameters.

In our work, we adopt a different approach from the literature for constructing low-delay erasure codes, where we do not optimize the performance of the code over sliding-window channels of any kind. Instead, we focus on constructing low-delay codes with linear field size that can also correct erasure patterns consisting of both burst and multiple arbitrary erasures occurring simultaneously in the same window. Moreover, our code design is focused on minimizing the packet loss probability by allowing the recovery of a maximal number of erased packets in scenarios where the erasure pattern is beyond the error correction capability of the code. The parameters of our code are $N,B,$ and $T$, where $T$ represents the maximum decoding delay and $(N,B)$ determines the erasure correction capability of the code. Our codes are explicit, systematic and can be constructed for all values of $N, B$, and $T$, with a linear field size $q\geq T$. We provide simulation results on the packet loss probability of our code over the GE channel and compare the results to existing constructions in the literature. The comparison shows that our code can outperform existing streaming codes in terms of both packet loss probability and maximum delay, for the same code rate and lower field size. An important implication of our results is that constructing rate-optimal streaming codes with zero error over theoretical channels, such as sliding-window channels, does not necessarily translate into codes with optimal performance over practical channels, such as the GE channel.

%This paper is organized as follows. In Section~\ref{sec:2}, we present the system model and define the sliding-window-erasure channel and the GE channel. We also explain an equivalence that was established in~\cite{Martinian2007, Kumar2020IT, Vajha} between low-delay block codes and convolutional codes which follows from the so-called diagonal and horizontal interleaving techniques. In Section~\ref{sec:4}, we present the novel code construction and discuss its properties. We provide the details of the encoding and decoding schemes of our code and give multiple encoding and decoding examples. We also provide simulations results that highlight the performance gain presented by this code with respect to existing constructions. In Section~\ref{sec:5}, we provide a theoretical analysis on the performance of our code and devise an optimization framework for choosing the optimal code parameters based on a theoretical bound on the packet loss probability that we derive .

\section{Preliminaries}
\label{sec:2}
\subsection{System Model}
\label{model}
In what follows, we discuss coding approaches for tackling the low-delay packet transmission problem. Convolutional codes are typically used for the low-delay packet transmission setting, and the delay is measured in terms of the time units that one needs to wait for before being able to decode an erased packet. These codes can be constructed in multiple ways but they share the basic idea of introducing redundant symbols or packets through time. The redundancy is generated by encoding the information symbols or packets causally. More specifically, the parity symbols generated at time $i$ can only depend on the information symbols at times $j\leq i$. We will be discussing two main approaches for constructing such convolutional codes that differ in terms of the allocation of redundancy across time instances. We will focus on systematic code constructions since the systematic property ensures that unerased packets  are recovered with zero delay.

The first known approach~\cite{Martinian2004} for designing low-delay codes is the following. Consider the setting where at each time~$i$, a source generates an information packet denoted by \mbox{$\mathbf{u}[i]=(u_1[i],u_2[i],\ldots,u_{k}[i])\in \mathbb{F}_q^k$} consisting of $k$ symbols over $\mathbb{F}_q$. At time $i$, a systematic encoder maps the information packet $\mathbf{u}[i]$ into an $n$-symbol coded packet denoted by \begin{equation}
\label{de}
\mathbf{x}[i]=(u_1[i],\ldots,u_{k}[i],p_1[i],\ldots,p_{n-k}[i])\in \mathbb{F}_q^n,
\end{equation}
where the parity symbols $p_1[i],\ldots,p_{n-k}[i]$ are generated as a function of $\mathbf{u}[j]$ with $j\leq i$. The rate of the code is $R=k/n$. Our goal is to design the code such that it allows decoding an erased packet under a maximum delay constraint $T$, where $T$ is a code parameter. Therefore, in the case where $\mathbf{x}[i]$ is erased, the underlying code should allow recovering the information packet $\mathbf{u}[i]$ by the time the packet $\mathbf{x}[i+T]$ is received. 

In the previous description, a delay-constrained code with rate $R=k/n$ is constructed by introducing $n-k$ redundant parity symbols at each time $i$. Alternatively, a second known approach~\cite{VajhaITW, Vajha} for constructing a code with the same delay properties and same rate $R=k/n$, is to transmit $k$ information packets, followed by $n-k$ parity packets, in $n$ consecutive time instances. More specifically, the first $n$ transmitted packets would be given by
\begin{equation}
\label{he}
  \mathbf{x}[i] = \left\{\def\arraystretch{1.2}%
  \begin{array}{@{}c@{\quad}l@{}}
    \mathbf{u}[i] \in \mathbb{F}_q^k, & \text{for $i=1,\ldots,k$},\\
	\mathbf{p}[i] \in \mathbb{F}_q^k, & \text{for $i=k+1,\ldots,n$},\\
  \end{array}\right.
\end{equation}
where $\mathbf{p}[k+1], \ldots, \mathbf{p}[n]$ are $n-k$ parity packets of length $k$ that are generated as a function of the $k$ preceding information packets $\mathbf{u}[1],\ldots,\mathbf{u}[k]$. This process can be repeated for every group of $k$ information packets. %The overall rate of the code is given by the ratio of the total number of information packets to the total number of transmitted packets, which is $R=k/n$.

\subsection{Diagonal and Horizontal Interleaving}
\label{interleaving}
Next, we present two known interleaving techniques for transforming a low-delay block code into a convolutional one with equivalent code properties. The importance of these interleaving techniques is that they reduce our problem to designing classical block codes that satisfy certain properties. 

Consider an $(n,k)$ block code $\mathcal{C}\subseteq \mathbb{F}_q^n$. Suppose that a codeword $\mathbf{x}=(x_1,\ldots,x_n)\in \mathcal{C}$ is affected by erasures resulting in a string $\mathbf{y}=(y_1,\ldots,y_n)$, where $y_i\in \mathbb{F}_q \cup \{\star\}$. We say that an erased codeword symbol $x_i$ is decoded with maximum delay $T$, if this symbol can be recovered using $(y_1,y_2,\ldots,y_{i+T^*})$, where $T^*=\min \{T,n-i\}$.
The two interleaving techniques that we explain next, called diagonal and horizontal interleaving, allow transforming any block code $\mathcal{C}$ into a convolutional code that has the same code rate and equivalent decoding delay properties.

The diagonal interleaving technique~\cite{Martinian2007} is applied in the setting corresponding to~\eqref{de}. In diagonal interleaving, the convolutional code is obtained from its block code counterpart~$\mathcal{C}$ by positioning the codeword symbols diagonally across time instances, i.e., for all $i$, the symbols of the transmitted packet $\mathbf{x}[i]=(x_1[i],\ldots,x_n[i])\in \mathbb{F}_q^n$ satisfy the following property
 $\left(x_1[i],x_2[i+1],x_3[i+2],\ldots,x_n[i+n-1]\right)\in \mathcal{C}.$  
 
 The horizontal interleaving technique~\cite{VajhaITW,Vajha} is applied in the setting corresponding to~\eqref{he}. Let $\mathcal{I}\triangleq \{1+mn~|~m\in \mathbb{N}\}$. In horizontal interleaving, each transmitted packet $\mathbf{x}[i]$ is of length~$k$, such that for $i\in \mathcal{I}$ we have 
\[
  \mathbf{x}[i+a-1] = \left\{\def\arraystretch{1.2}%
  \begin{array}{@{}c@{\quad}l@{}}
    \mathbf{u}[i+a-1]  \in \mathbb{F}_q^k, &  a=1,2,\ldots,k,\\
	\mathbf{p}^{a-k+1}[i+a-1] \in \mathbb{F}_q^k, & a=k+1,\ldots,n,\\
  \end{array}\right.
\]
where $\mathbf{p}^1[i+k],\mathbf{p}^2[i+k+1], \ldots,\mathbf{p}^{n-k}[i+n-1]$ represent the $n-k$ parity packets generated by horizontally encoding the symbols of $\mathbf{u}[i], \mathbf{u}[i+1],\ldots,\mathbf{u}[i+k-1]$ using the underlying systematic code $\mathcal{C}$. Namely, for $i\in \mathcal{I}$ and $j=1,\ldots,k$, 
$$\left( u_j[i],\ldots,u_j[i+k-1], p^1_j[i+k],\ldots,p^{n-k}_j[i+n-1]\right) \in \mathcal{C}.$$

One can easily show that the aforementioned interleaving techniques result in a convolutional code that has equivalent decoding properties as the block code $\mathcal{C}$~\cite{Martinian2007, Kumar2020IT, VajhaITW, Vajha}. For more details on these interleaving techniques, we refer interested readers to~\cite{Martinian2007, Vajha} and references therein.

\section{Code Construction}
\label{cons}
%In this section, we present our code construction and discuss its erasure correction capabilities.
\subsection{Encoding}
\label{encoding}
Suppose that we are given a target maximum delay $T$, and we choose certain code design parameters $B\geq 0$ and $N\geq 0$. Let \mbox{$\mathcal{C}\subseteq \mathbb{F}_q^n$} be an $(n,k)$ block code with $k=T-N$ and $n=T+B$, such that $B+N\leq T$ and $q\geq T$. Let $\mathbf{G}$ be the $k\times n$ generator matrix of $\mathcal{C}$. Let $\mathbf{u}=(u_1,\ldots, u_k)\in \mathbb{F}_q^k$ be the information message and $\mathbf{x}\in \mathbb{F}_q^n$ be the codeword. Since the code is systematic, we have $\mathbf{x}=\mathbf{u}\mathbf{G}=(u_1,\ldots, u_k, p_1, \ldots, p_{n-k}),$ where $p_1, p_2, \ldots, p_{n-k}$ represent the $n-k=B+N$ parity symbols. Next, we present our construction by giving the expressions of the parity symbols $p_1, \ldots, p_{B+N}$, as a function of the information symbols $u_1,\ldots, u_k$. 

\begin{enumerate}[leftmargin=*]
\item The first $N$ parity symbols $p_{1},p_{2},\ldots,p_{N}$ are generated using a systematic MDS code over $\mathbb{F}_q$. For explicitness, the systematic Reed-Solomon (RS) encoder based on the Cauchy matrix construction is used. %The punctured code obtained by only considering the first $k+N$ symbols of each codeword, i.e., $(u_1,\ldots, u_k, p_{1},\ldots,p_{N})$, is a $(k+N,k)$ systematic Reed-Solomon code.

\item The $B$ parity symbols $p_{N+1},p_{N+2},\ldots,p_{N+B},$ are generated as interleaved parity checks with an interleaving factor of $B$, such that $\forall i \in \{1,2,\ldots,B\}$,
\begin{equation}
\label{p2}
p_{N+i}=u_i+u_{i+B}+\ldots+u_{i+(q-1)B}+ \mathds{1}_{\{i\leq r\}} u_{k-r+i},
\end{equation}
\end{enumerate}
where $k=qB+r$ and $r=k\Mod B$. The final step of the construction is to transform this block code into a convolutional one by applying either diagonal or horizontal interleaving, as explained in Section~\ref{interleaving}.

\subsection{Decoding}
\label{decoding}
To explain the decoding scheme, we start by examining the simple case when the total number of erasures in $(x_1,\ldots,x_{k+N})$ is $\leq N$. Notice that the punctured code obtained by only considering the first $k+N$ symbols of each codeword, is an $(k+N,k)$ systematic RS code. Hence, if the total number of erasures in the first $k+N$ symbols is $\leq N$, we apply RS decoding to recover these erasures. In this case, it follows from the code design that the erased symbols can be recovered with a delay that is $<T$.

Next, we examine the case when the total number of erasures in $(x_1,\ldots,x_{k+N})$ is $> N$. In this case, we are interested in decoding as many information symbols (packets) as possible while respecting the maximum delay constraint $T$. It follows from the code design that the first $B-1$ information symbols $u_1,u_2,\ldots,u_{B-1}$ fall within the maximum decoding delay constraint $T$ since these symbols need to be decoded before the end of the code. Whereas the delay constraint for the remaining information symbols $u_{B},u_{B+1},\ldots,u_{k}$ is inactive because the constraint falls right at the end or past the end of the code. We refer to the symbols $u_1,\ldots,u_{B-1}$ as urgent symbols, and to $u_{B},\ldots,u_{k}$ as non-urgent symbols. 

We start our decoding process by decoding the urgent symbols $u_1,\ldots,u_{B-1}$ symbol-by-symbol from left to right. Suppose that $u_1$ is erased, we attempt to decode $u_1$ using the interleaved parity check $p_{N+1}$. If $u_1$ is the only summand that is erased in the interleaved parity check $p_{N+1}$, then $u_1$ can be successfully recovered by a simple addition/subtraction operation. Otherwise, $u_1$ is considered to be lost because it could not be decoded within its delay constraint. Then, similarly, we proceed to decode the next erased urgent symbol. If at any point of this symbol-by-symbol decoding process, the number of erased symbols in $(x_1,\ldots,x_{k+N})$ becomes $\leq N$, we use the punctured RS code to decode all symbols, and the decoding is subsequently terminated. In some scenarios, an urgent symbol may not be decodable with delay $\leq T$, but could be decoded at later stages of the decoding process with delay $>T$. In such scenarios, we still consider the urgent symbol to be lost since its recovery was not successful within the maximum decoding delay constraint $T$.

As for the non-urgent symbols $u_{B},\ldots,u_{k}$, the general decoding approach for these symbols consists of two phases. In the first phase, we decode as many symbols as possible using the interleaved parity checks $p_{N+1},\ldots,p_{N+B}$, given in~\eqref{p2}. In the second phase, we decode the remaining symbols that were not recovered in phase 1 by using the RS parities $p_1,\ldots,p_N$. The worst-case overall decoding complexity is $O(T^2)$~\cite{berlekamp}.

\subsection{Erasure Correction Capability}
%\hl{Let $\mathbf{e}_i\in \{0,1\}^k$ be the $i^{th}$ canonical basis vector of $\{0,1\}^k$ and $\Tilde{G}(E)$ be the generator matrix punctured at the non-zero entries of a given erasure pattern $E\in{0,1}^n$. In general, for a given erasure pattern $E$, an erased information symbol $u_i$, $i=1,\ldots,k$, can be decoded if $\mathbf{e}_i$ is in the column span of $\Tilde{G}$ function of erasure pattern and $T$. To obtain a better understanding of the code, we detail in this section some of the cases where our code can decode successfully under the maximum delay constraint $T$.}

In this section, we discuss the erasure correction capability of the code in terms of the code parameters $N,B,$ and $T$. Namely, we detail some of the cases where {\em all} erasures can be decoded, and we present a general condition under which the partial recovery of {\em some} erasures is possible.

%we show that the code can correct all erasure patterns consisting of $N'\leq N$ arbitrary erasures, and most erasure patterns consisting of $B'\leq B$ burst erasures, while respecting the maximum decoding delay constraint $T$. Furthermore, we detail some of the cases where the code can correct erasure patterns consisting of a combination of \mbox{$B'\leq B$} burst erasures in addition to $N'\leq N$ arbitrary erasures.

{\em Case 1:} Suppose that the erasure pattern consists of \mbox{$N'\leq N$} arbitrary erasures and no other erasures are present. In this case, all erased information symbols can be decoded using the punctured RS code obtained by only considering the first $k+N$ codeword symbols. Furthermore, it follows from the construction that the corresponding decoding delay is $<T$. 

{\em Case 2:} Suppose that the erasure pattern consists of a burst of size $B'\leq B$ and no other erasures are present. In this case, we show that, under certain conditions, all the erased symbols can be decoded using the interleaved parity checks $p_{N+1},\ldots,p_{N+B}$. We focus our discussion on the case where all the symbols are erased in a given burst since it is the most challenging one. Recall that $k=qB+r$, with $r=k\Mod B$. For $i\in \{1,2,\ldots,B\}$, if $r=0$, let $S_i= \emptyset$; otherwise, let $S_i= \{u_{k-r+i}\}$, and define the sets 
$ U_i\triangleq \{u_i, u_{i+B}, u_{i+2B}, \ldots, u_{i+(q-1)B}\} \cup S_i, $
which partition the set of information symbols $\{u_1,\ldots,u_k\}$. Notice that $U_i$ contains all the information symbols that appear as summands in the interleaved parity check $p_{N+i}$.  Assume that the burst erases the systematic symbols $u_{i'},u_{i'+1},\ldots,u_{i'+B'-1}$, with $i'\in \{1,\ldots,k-B'+1\}$. It follows from the construction of the interleaved parity checks that the \mbox{$B'\leq B$} consecutive message symbols $u_{i'},\ldots,u_{i'+B'-1}$ belong to different partitions $U_i$. Therefore, any burst of size $B'\leq B$ can erase at most one of the summands in $p_{N+1},p_{N+2},\ldots,p_{N+B}$. Hence, the erased symbols $u_{i'},\ldots,u_{i'+B'-1}$ can be easily decoded by a simple subtraction/addition operation applied on the parity checks.

The argument above covers the case where the burst affects only the systematic symbols. Since the code is systematic, the case where the burst affects only the parity symbols is trivial. For this reason, we examine next the case where the burst affects some systematic and parity symbols simultaneously. Suppose that a systematic symbol $u_i$, $i\in \{1,\ldots,k\}$, appears as a summand in the interleaved parity check $p_j$, $j\in \{N+1,\ldots,N+B\}$. If $B \vert k$, i.e., $r=0$, it follows from the definition of the interleaved parity checks and their positioning in the codeword that any burst of size $B'\leq B$ cannot erase $u_i$ and $p_j$ simultaneously. Namely, if $r=0$, the construction guarantees the availability of every systematic symbol in at least one coded symbol after any $B'\leq B$ burst erasures. Hence, all erased information symbols can be decoded using $p_{N+1},\ldots,p_{N+B}$. Also, if $B\nmid k$, i.e., $r>0$, one can easily show that any $B'$ burst erasures can be decoded if $B'\leq N+r$. Furthermore, it follows from the construction that the decoding delay is $\leq T$ for all recovered symbols.
 
{\em Case 3:} Suppose that the erasure pattern consists of a burst of size $B'\leq B$ in addition to $1\leq N'\leq N$ arbitrary erasures. We will focus on the case where $B'=B>N$. Consider the case where the burst affects the non-urgent symbols $u_{B},\ldots,u_{k}$. In this case, the following general rule applies. The code can correct $B$ burst erasures in addition to $N'\leq N$ arbitrary erasures if the number of erasures to be decoded in phase~2 of the decoding process (Section~\ref{decoding}) is $N_2<N$, where $N_2$ denotes the number of erasures to be decoded in phase~2. To understand this, recall that any burst of size $B$ can erase only one of the summands in $p_{N+1},\ldots,p_{N+B}$. Hence, if $N'$ arbitrary erasures occur in addition to the $B$ burst erasures, then at least $B-N'$ out of the $B$ interleaved parity checks $p_{N+1},\ldots,p_{N+B}$ will have only one erased summand (that can be recovered by simple addition/subtraction operation). Therefore, by the end of decoding phase~1, at least $B-N'$ out of the $B+N'$ erasures can be successfully decoded. Consequently, the number of erasures to be decoded in phase~2 satisfies $N_2 \leq B+N'-(B-N')=2N'$. Using the $(k+N,k)$ punctured code 
systematic RS code, up to $N$ erasures can always be decoded in phase 2. Notice that if $N'\leq N/2$, then $N_2\leq N$, and hence, all erased symbols can be decoded successfully. Whereas if $N'>N/2$, then some erasure patterns, but not all, can be corrected depending on the locations of the $N'$ arbitrary erasures. We omit the discussion of more cases due to space limitations. %Suppose that the burst erases the symbols $u_{B+1},\ldots,u_{2B}$, and all the $N'$ arbitrary erasures affect symbols belonging to the set $U_1$ defined in~\eqref{partition}. In this case, the $B-1$ information symbols $u_{B+2}, \ldots, u_{2B}$ can be decoded in phase~1 using the interleaved parity checks $p_{N+2},\ldots,p_{N+B}$. Hence, the number of erasures to be decoded in phase~2 is $N'+1$. Thus, the arbitrary erasures can be corrected in this case using the punctured Reed-Solomon code if $N'\leq N-1$.

Furthermore, a key feature of our code is that it allows the recovery of a large number of erased symbols in many cases where the erasure pattern is not correctable. Note that this is contrary to codes such as $(n,k)$ systematic RS codes, where not a single erasure can be recovered when the total number of erasures is $> n-k$. In practice, this decoding feature translates into a low packet loss probability as we show in Section~\ref{simulnew}. The general condition under which an  information symbol can be recovered with maximum delay $T$ is given in Proposition~\ref{prop1}. %This partial recovery feature is one of the main reasons that allows our to code to outperform existing streaming codes as we show in Section~\ref{simulnew}.

\begin{proposition}
\label{prop1}
Let $\bm{\varepsilon}\in\{0,1\}^{n}$ denote an erasure pattern, where $\varepsilon_j=1$, $j=1,\ldots,n$, indicates that the codeword symbol $x_i$ is erased. Let $\mathbf{e}_i\in \{0,1\}^k$, $i=1,\ldots,k$, be the $i^{th}$ canonical basis vector, and $\Tilde{\mathbf{G}}(\bm{\varepsilon}_1^{j})\in \mathbb{F}_q^{k\times n^*}$ be the generator matrix of the code $\mathcal{C}$ (Section~\ref{cons}) punctured column-wise at the indices of $\bm{\varepsilon}_1^{j}=(\varepsilon_1,\ldots,\varepsilon_j)$ whose entries are zero, with $n^*=j-\sum_{i=1}^j \varepsilon_i$. For a given $\bm{\varepsilon}$, an information symbol $u_i$ can be decoded with maximum delay $T$ if and only if $\mathbf{e}_i$ is in the column span of $\Tilde{G}(\bm{\varepsilon}_1^{i+T^*})$, with $T^*=\min \{T,n-i\}$.
\end{proposition}

\section{Simulation Results}
\label{simulnew}
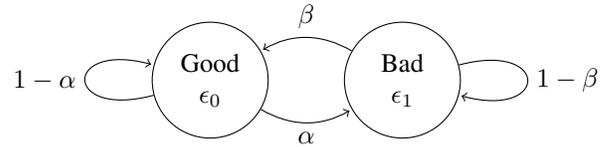
\begin{figure}[h!]
%\vspace{0.2cm}
\centering
\begin{tikzpicture}[scale=0.2]
        % Draw the states
        \node[state, text width=1cm, align=center]             (s) {Good $\epsilon_0$};
        \node[state,  text width=1cm, align=center, right=of s] (r) {Bad \\ $\epsilon_1$};

        % Connect the states with arrows
        \draw[every loop]
            (r) edge[bend right, auto=right] node {$\beta$} (s)
             (s) edge[bend right, auto=right] node {$\alpha$} (r)
            (s) edge[loop left] node {$1-\alpha$} (s)
	(r) edge[loop right] node {$1-\beta$} (r);
    \end{tikzpicture}
\caption{{\small 2-state Markov chain illustration of the GE channel. The erasure probabilities in the good and bad states are denoted by $\epsilon_0$ and $\epsilon_1$, respectively. The transition probabilities between the good and bad states are denoted by $\alpha$ and $\beta$.}}
\label{fig2}
\end{figure}

In this section, we compare the packet loss probability (PLP) of our proposed code with other existing constructions over the GE channel (Figure~\ref{fig2}). The empirical PLP is computed as the ratio of the total number of lost information packets to the total number of transmitted information packets. The properties of the simulated codes are shown in Table~\ref{tabn1}. Due to space limitations, we only show results for the case of \mbox{$\alpha=5\times 10^{-3}$}, $\beta=0.45$, $\epsilon_1=1$, and varying $\epsilon_0=\epsilon$. %We show results corresponding to the horizontal interleaving technique (Section \ref{interleaving}), other results for the diagonal interleaving method are available in~\cite{extended}.  In addition to our code construction, we study the performance of the codes in table~\ref{tabn1}.
%following codes:  \begin{enumerate*}[label={\textit{(\roman*)}}] \item Horizontally interleaved $(n,k)$ MDS code that can correct any $N=B$ erasures, where $n=T+1$, $k=n-N$ and $q=O(T)$; \item~Horizontally interleaved $(n,k)$ Martinian-Trott code~\cite{Martinian2007} that can correct a burst of maximum size $B$, with $k=T$, $n=k+B$, and $q=O(T)$; \item Horizontally interleaved $(n,k)$ Domanovitz {\em et al.}~\cite{Domanovitz2020} code that can correct a burst of maximum size $B$ {\em or} up to $N$ arbitrary erasures following the $C(N,B,W=T+1)$ channel model (Section \ref{GE}), where $k=T-N+1$, $n=k+B$, and $q=O(T^2)$. \end{enumerate*} Recall that our $(n=k+B+N,k=T-N)$ horizontally interleaved code has parameters $B$ and $N$ which denote the number of interleaved parity checks and Reed-Solomon parities, respectively. Also, the field size requirement is $q\geq T$. The properties of the codes are given in Table~\ref{tabn1}. 

\begin{table}[h!]
\centering
\begin{tabular}{|c|c|c|c|c|}
\hline
& $T$ & $(n,k)$ & Rate & $(N,B)$  \\ \hline
MDS Code  & $15$     & $(16,8)$ & $0.5$ &$(8,8)$  \\ \hline
Martinian-Trott Code~\cite{Martinian2007}   & $15$     &  $(30,15)$ & $0.5$ &$(1,16)$  \\ \hline
Domanovitz et al. Code~\cite{Domanovitz2020}   & $15$     &  $(24,12)$ & $0.5$ &$(4,12)$\\ \hline
New Code 1   & $15$     &  $(22,11)$ & $0.5$ &$(4,7)$ \\ \hline
New Code 2   & $14$     &  $(20,10)$ & $0.5$ &$(4,6)$ \\ \hline
%New Code 3 &$(9,1)$  & $10$     &  $(19,9)$ & $0.47$  \\ \hline
%New Code 4 &$(7,1)$  & $8$     &  $(15,7)$ & $0.47$  \\ %\hline
\end{tabular}
\caption{{\small The properties of the codes simulated in Figure~\ref{newf1}.}} 
\label{tabn1}
\end{table}

\begin{figure}[h!]
\vspace{-0.5cm}
\centering
\includegraphics[width=0.47\textwidth]{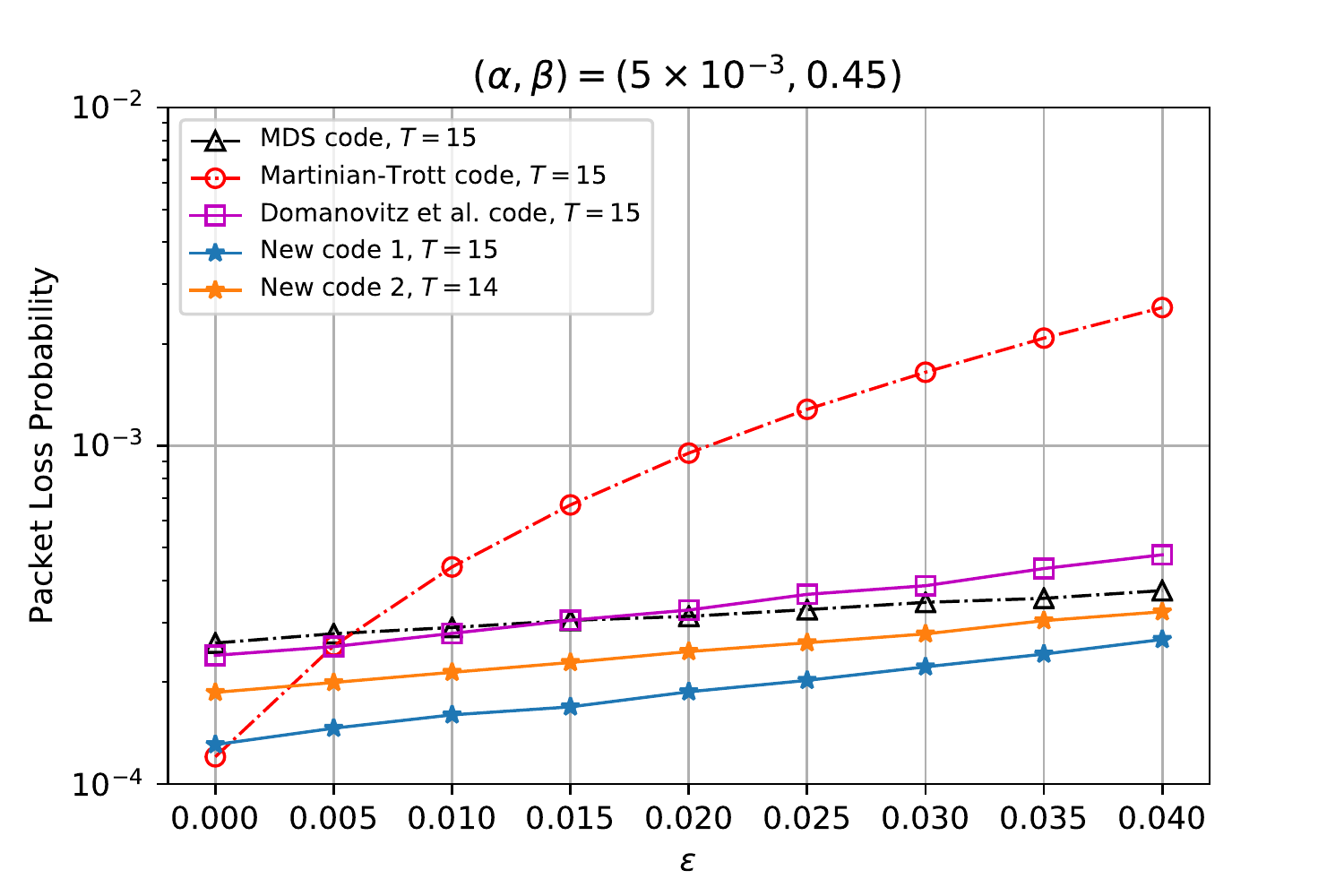}
\caption{{\small PLP of the codes in Table~\ref{tabn1} over the GE channel with $\alpha=$5e\text{-}3$, \beta=0.45$, and varying $\epsilon$. The horizontal interleaving setting is considered (Section \ref{interleaving}) and the channel length is set to $10^7$.}}
\vspace{-0.7cm}
\label{newf1}
\end{figure}

\begin{comment}
\begin{figure}[h!]
\centering
\includegraphics[width=0.48\textwidth]{Figures/Pall1.pdf}
\caption{Comparing the packet loss probability of different code construction over the GE channel while varying $\epsilon$. The simulation results are for $\alpha=5\times 10^{-4}, \beta=0.5$. The properties of the compared codes are given in Table~\ref{tabn1}. The channel length is set to $10^7$.}
\label{newf2}
\end{figure}
\end{comment}

The results in Figure~\ref{newf1} show that for $\epsilon=0$, the Martinian-Trott code, which is optimized for correcting bursts, achieves the best performance. However, its performance rapidly deteriorates as $\epsilon$ increases. For $0<\epsilon\leq 0.04$, our code with $(N=4, B=7, T=15)$ achieves the lowest PLP among all constructions. Also, for $\epsilon\geq 0.005$, our code with $(N=4, B=6, T=14)$ presents a twofold gain since both the PLP and the maximum delay $T$ are lower than that of other constructions. Here it is noteworthy to also mention that these constructions require different field sizes. For $R=0.5$, the Martinian-Trott code can be constructed in $GF(2)$, the MDS code and our code require a linear field size $q\geq T$, and the Domanovitz code requires a quadratic field size $q=O(T^2)$.

%Note that for a given code rate, there are multiple ways to construct our code and the code by Domanovitz {\em et al.}~\cite{Domanovitz2020} depending on the choice of the code parameters $B$ and $N$. In Figure~\ref{newf1}, we only show the constructions that gave the lowest overall empirical packet loss probability among all possible choices of $B$ and $N$ with $R=0.5$ and the given values of $T$. In the extended version of this paper~\cite{extended}, we provide a theoretical probabilistic analysis of the performance of our code over the GE channel, which allows us to determine the optimal choice of the code parameters $B$ and $N$ for any set of system and channel parameters.  

In conclusion, the results demonstrate that optimizing code constructions over theoretical sliding-window channels does not necessarily give the best performance over practical channels such as the GE channel. Namely, the results in Figure~\ref{newf1} show that with the same code rate and lower field size, our code can outperform existing rate-optimal streaming codes in terms of both packet loss probability and maximum decoding delay. As part of future work, our goal is to derive a theoretical bound on the PLP of our code over the GE channel.

\bibliographystyle{ieeetr}

\bibliography{Refs4}
\end{document}